\documentclass[twocolumn,aps,prl,showpacs,preprintnumbers,amsmath,amssymb, superscriptaddress]{revtex4}

\usepackage{graphicx}
\usepackage{dcolumn}
\usepackage{bm}

\begin{document}


\title{Critical Behavior and Extended States in 2D and 3D Systems with Gas-like Disorder}


\author{D. J. Priour, Jr}
\affiliation{Department of Science, Kansas City Kansas Community College, Kansas City, Kansas 66112, USA}


\date{\today}

\begin{abstract}
With a tight binding treatment we examine amorphous conductors with 
gas-like disorder, or no correlations among the site positions.  We      
consider an exponentially decaying hopping integral with range $l$, 
and the Inverse Participation Ratio (IPR) is used to characterize 
carrier wave functions with respect to localization.  With the  
aid of two complementary finite size scaling techniques to extrapolate 
to the bulk limit (both methods exploit critical behavior in 
different ways to find the boundary between domains of extended 
and localized wave functions) which nevertheless yield identical 
results, we obtain phase diagrams showing regions where states  
are extended and domains of localized states.  In the 
2D case, states are localized below a threshold length scale $l_{c}$ on the 
order of the interparticle separation $\rho^{-1/2}$ with a finite 
fraction of states extended for $l > l_{c}$.  For $D = 3$, the 
extended phase is flanked by regions of localized states and bounded 
by two mobility edges.  The swath of extended states, broad for 
$l \sim 1$, becomes narrower with decreasing $l$, though persisting with
finite width even for $l <  \frac{1}{5}\rho^{-1/3}$.  Mobility edges are 
interpreted as lines of critical points, and we calculate the corresponding 
critical exponents.
\end{abstract}

\maketitle


The Anderson model~\cite{Anderson}, with disorder manifested as random perturbations in the site potential 
(i.e. diagonal disorder) has provided a way to examine localization effects in 
the context of a tight binding treatment.  However, off-diagonal disorder where
there are stochastic variations in the 
tunneling matrix elements also is germane to situations examined in experiment~\cite{Edwards} and theoretical
studies~\cite{Dyson,Gade,Parshin,Cerovski,Takahashi} which have also included 
calculations of the energy density of states (DOS)~\cite{Logan1,Logan2,Logan3} and aspects of 
localization for 3D geometries~\cite{Ching,Blumen,Logan4,Krich,Xiong}.  
Our aim is to determine conditions for the existence of extended states 
(system spanning wave functions amenable to charge transport) in the presence of very 
strong disorder where there is no correlation among site positions, a circumstance 
known as gas-like disorder~\cite{Ziman} and also termed topological disorder due to 
the absence of a specific underlying lattice geometry.

Setting the constant diagonal terms to zero, we consider the tight binding Hamiltonian
\begin{align}
{\mathcal H} = -\frac{1}{2} t_{0} \sum_{i=1}^{N} \sum_{j \neq i} V(r_{ij}) (\hat{c}_{i}^{\dagger} \hat{c}_{j} + 
\hat{c}_{i} \hat{c}_{j}^{\dagger}) 
\label{Eq:eq1}
\end{align}
where the sum over the index ``$i$'' ranges over the $N$ particles contained in the simulation 
volume, we take the hopping parameter $t_{0}$ to be 3.0 electron volts, and the factor of ``1/2'' 
compensates for multiple counting.  The  
operators $\hat{c}^{\dagger}$ and $\hat{c}$ create and destroy occupied electronic orbitals at 
sites indicated by the subscript.  For the hopping integral, we use an exponential 
dependence $V(r_{ij}) = e^{-\gamma r_{ij}/s}$, 
where $r_{ij}$ is the separation between sites $i$ and $j$,
$s = \rho^{-1/D}$ is the typical distance between neighboring sites, $\rho$ 
is the volume density of sites, $D$ is the dimensionality of the system, and  
$\gamma$ is a dimensionless parameter.   
Since the length scale for the decay of the tunneling matrix element is $l = s/\gamma$, large/small $\gamma$ 
values correspond to decay lengths small/large in relation to the typical interatomic separation.
For the sake of convenience, we rescale coordinates such that $\rho = 1$, with a simple 
inverse relationship  $l = \gamma^{-1}$,    
and $V(r_{ij}) = e^{-\gamma r_{ij}}$.
Although the matrix element $V(r_{ij})$ is finite in range by virtue of the exponential decay,
we nevertheless take into consideration hopping among all pairs of orbitals contained in the system of 
$N$ sites with only a negligible increase in computational burden
(i.e. a contribution on the order of $N^{2}$) relative to direct 
diagonalization used to calculate carrier states, which scales as $N^{3}$. 

With positional order (either locally or 
globally) not preserved in gas-like amorphous materials, 
the disorder strength is nonetheless characterized in a sense by 
the hopping integral decay scale $l$.  For large $l$ (small $\gamma$), the 
comparatively long-ranged tunneling connects sites to many neighbors, effectively  
averaging  over hopping rates to and from many sites and mitigating the 
effect of strong disorder.  However, if $l \ll 1$ (large $\gamma$), 
tunneling chiefly involves nearest neighbors.  
Hence, disorder is an a sense amplified for $\gamma \gg 1$ and muted if $\gamma \ll 1$,
and one proceeds with the intuition that extended states are more likely to exist
for larger $l$ than for smaller $l$ values.

For 2D systems, we find a threshold $l_{c}$ where all states are localized for $l > l_{c} \sim 1$, 
with a finite portion of the wave functions extended for $l < l_{c}$.
In the 3D case,
a meta-study~\cite{Edwards} of a large body of experiments in disparate systems corresponding to a broad 
range of $\gamma$ values suggests 
a termination of conducting behavior for $\gamma > 3.8$.   
Nevertheless, though we find the portion of extended states to diminish monotonically in 
$\gamma$, we do not find an abrupt termination of the extended region for any
particular value of $\gamma$, in contrast to $D = 2$ where extended states do not 
exist if $l$ is smaller than the interparticle separation.  
We construct phase diagrams showing a region of extended states for 
intermediate energies bounded by two mobility edges; for a specific $\gamma$ value, 
the latter are critical points marking a second order transition from the 
localized to extended phase at the lower edge and from extended to localized states at the upper 
edge.

To implement the numerical calculations, we examine a $L^{D}$ 
supercell using periodic boundary conditions to mitigate 
finite size effects. A range of $L$ values (i.e. as high as $L = 110$ for $D = 2$ and 
$L = 22$ for $D = 3$) are used to perform 
finite size scaling to determine if electronic states are localized 
or extended in the bulk limit.  The Inverse Participation Ratio,
$Y_{2} = \sum_{i=1}^{N} |\psi_{i} |^{4}/(\sum_{i=1}^{N} | \psi_{i} |^{2})^{2}$   
shows distinct behavior depending on whether the wave function is confined to 
a small volume (larger $Y_{2}$) or spread out over a larger region (smaller $Y_{2}$), and hence more extended in 
character.
With a crystal lattice absent, one operates in the grand canonical ensemble, 
and the number of sites $N$ varies about $\langle N \rangle = \rho L^{D}$ from sample to sample.
In generating realizations of disorder,
we account for fluctuations in the numbers of particles 
in an unbiased way using a technique   
described elsewhere~\cite{priour}.  To obtain adequate statistics,
$10^{5}$ wave functions are retained for each combination of $L$ and 
$\gamma$. 

The random character of disorder precludes the study of the evolution of 
individual states with increasing $L$, and instead aggregates of wave 
functions must be examined.  With carrier states parameterized by energy 
eigenvalues, one could in principle create partitions 
of width $\delta E$
centered about uniformly spaced energies.  However, since the energy density 
of states is sharply peaked~\cite{priour}, energy channels far from the central 
peak suffer from poor statistics.  We ensure uniform statistics by 
using normalized rank $\tilde{r} = r/M$ (ranging from 0 to 1) to label carrier states,
where $M$ is the total number of states and $r$ is the global eigenvalue rank within
the large aggregate.
For $n = 100$ partitions  
of the $\tilde{r}$ domain, the channel width $\delta \tilde{r}$ 
is sufficiently small that systematic admixture effects from neighboring channels play a 
minor role, while still providing sufficient statistics for analysis, and 
parallel calculations for $n = 50$ and $n = 200$ yield results in quantitative 
agreement with the $n = 100$ scheme.

We calculate the channel averaged IPR 
in the thermodynamic limit as an avenue for the 
identification of 
mobility edges.  For moderate to large $L$, we use a power law     
formula $Y_{2}(L) = Y_{2}^{0} + A_{1} L^{-\beta} + 
A_{2} L^{-\delta}$ where $Y_{2}^{0}$ is the participation ratio in 
the bulk limit, $\beta$ and $\delta$ are leading and next-to-leading
order exponents, and 
$A_{1}$ and $A_{2}$ are amplitudes.  The parameters 
$Y_{2}^{0}$, $\beta$, $\delta$, $A_{1}$, and $A_{2}$
are fixed by a Monte Carlo nonlinear least squares fit to the 
channel averaged participation ratios.

\begin{figure}
\includegraphics[width=.45\textwidth]{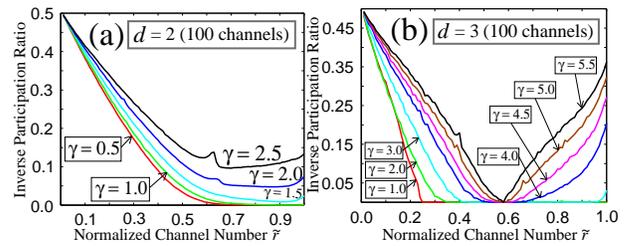}
\caption{\label{fig:Fig1} (Color Online) Extrapolated Participation 
Ratio 2D and 3D results appear in panel (a) and panel (b), respectively
for various $\gamma$ values.}
\end{figure}

Extrapolated participation ratios appear in Fig.~\ref{fig:Fig1}
The troughs in the bulk $Y_{2}$ curves shown in panel (a) of Fig.~\ref{fig:Fig1}
deepen with decreasing $\gamma$; while IPR minima for $\gamma > 1.0$ 
are manifestly nonzero, participation ratios do seem to descend to 
zero for less rapid decay rates.  Nevertheless, the gradual approach to the 
abscissa confounds a straightforward objective identification of a mobility edge for the 
2D systems we examine.

Irrespective of $\gamma$, the $D = 3$ extrapolated IPR curves in panel (b) Fig.~\ref{fig:Fig1} seem to plunge to 
zero and remain zero for a range of $\tilde{r}$ values.  However, the 
width of the interval 
where extended states seem to exist decreases with $\gamma$, with 
$Y_{2}$ appearing to vanish for most of the $\tilde{r}$ domain 
in the case $\gamma = 1.0$ while only 
briefly touching the abscissa for $\gamma = 5.5$. 
For $\gamma < 3$ the IPR falls to zero with no subsequent return to  
finite values.  On the other 
hand, for $\gamma \geq 3$, the interval of vanishing participation ratios is 
terminated for $\tilde{r} < 1$ as $Y_{2}$ abruptly rises to finite values.
Salient feature not present in the 2D results are derivative discontinuities 
where $Y_{2}$ falls to zero, descending to or rising from the horizontal
axis with a finite slope.  The clear delineation of the regions where the 
IPR vanishes favors the use of basins in the $Y_{2}$ curves 
as a way to identify the boundary between extended
and localized states in the 3D case.

\begin{figure}
\includegraphics[width=.40\textwidth]{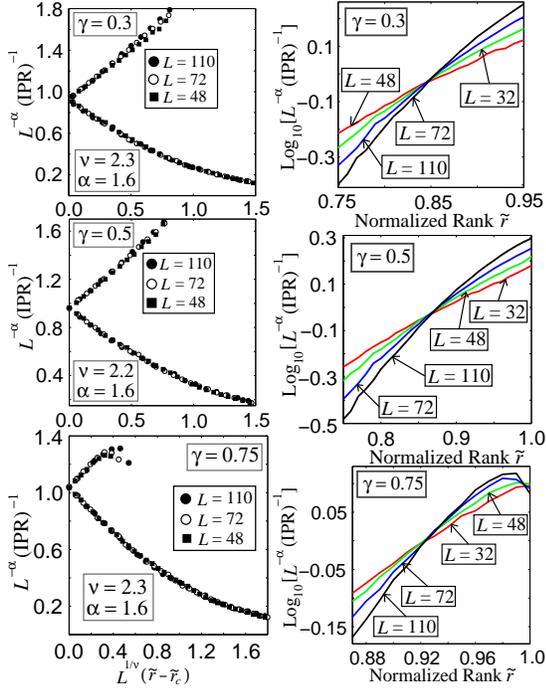}
\caption{\label{fig:Fig2} (Color Online) Data collapse plots for 
(top to bottom) $\gamma = 0.3$, $\gamma = 0.5$, and $\gamma = 0.75$ appear on the 
left with the corresponding $\phi_{2}$ intersections displayed on the right for $D = 2$.}
\end{figure}

\begin{figure}
\includegraphics[width=.45\textwidth]{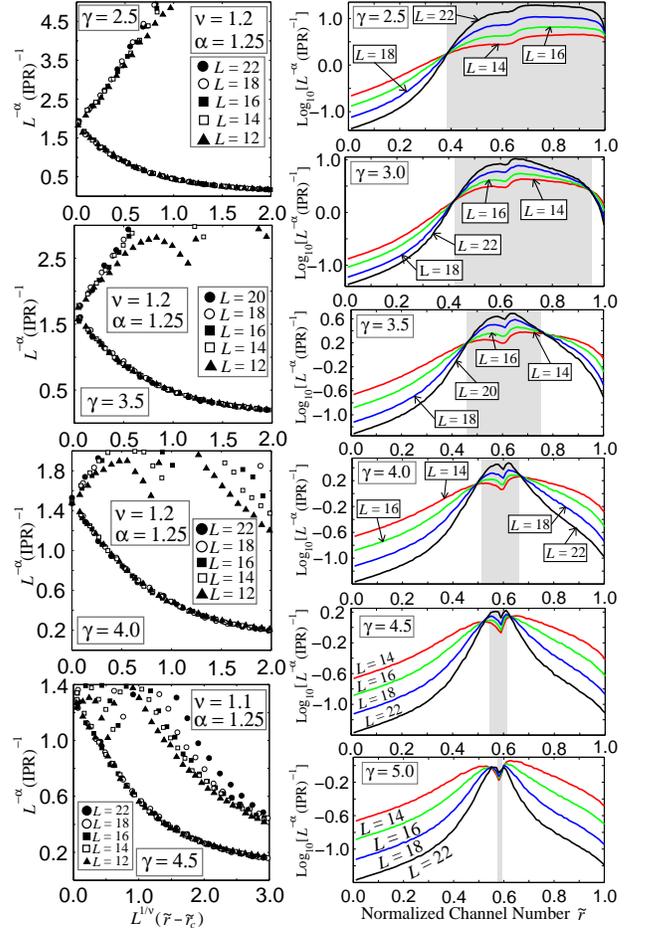}
\caption{\label{fig:Fig3} (Color Online) Data collapse plots for 
various $\gamma$ values ($\gamma$ increases from top to bottom); data collapse plots 
corresponding to the lower mobility edge appear on the left, 
with  $\phi_{2}$ intersection plots shown on the right for $D = 3$.}
\end{figure}

With a complementary analysis, we calculate phase portraits 
showing localized and extended states for
$D = 2$ and $D = 3$.  
As a consequence of critical behavior at the mobility edge,
one expects from single parameter finite size scaling theory a form 
$Y_{2} = L^{\alpha_{D}} F_{D}[L^{1/\nu_{D}} (\tilde{r} - \tilde{r}_{c})]$
near the boundary between
regions of extended and localized states, with $\alpha_{D}$ and $\nu_{D}$ being 
critical exponents ($\nu_{D}$ controls the singular behavior of the 
correlation length $\xi$).
Instead of trying to determine numerically where $Y_{2}$ falls to zero,
the mobility edge is identified in an objective fashion by locating  
intersections of 
$\phi_{2} \equiv Y_{2}(\tilde{r},L)^{-1}L^{-\alpha_{D}}$, where 
$\alpha_{D}$ is fixed by insisting that curves for different 
system sizes $L$ coincide, with $\tilde{r}_{c}$ determined 
by the location of the crossing.  A similar approach has been applied 
in theoretical studies of a 2D Anderson model with long-range correlated disorder~\cite{Santos}
and Anderson localization of phonons~\cite{Monthus} 
Data collapses, where $\phi_{2}$ points plotted versus 
$L^{1/\nu_{D}} (\tilde{r} - \tilde{r}_{c})$ for various $L$ 
coincide on a single curve also are associated 
with a critical point, and in the case of 
the lower mobility edge are sharp enough to
permit the determination the critical index $\nu_{D}$
by optimizing the data collapse.

Across the range of $\gamma$ examined, we find 
from the $\phi_{2}$ curve crossings that $\alpha_{\mathrm{2d}} = 1.6 \pm 0.1$    
for $D = 2$
and $\alpha_{\mathrm{3d}} = 1.25 \pm 0.15$ for $D = 3$.
In the 2D case, a mobility edge is identified only for $\gamma < 1$, and in 
the right panels of Fig.~\ref{fig:Fig2}, $\phi_{2}$ intersections for four 
distinct $L$ values are shown for $\gamma = \left \{ 0.3, 0.5, 0.75 \right \}$.
The left panels show good data collapses for $\nu_{\mathrm{2d}}(\gamma = 0.3) = 2.3 \pm 0.2$,
$\nu_{\mathrm{2d}}(\gamma = 0.5) = 2.2 \pm 0.2$, and $\nu_{\mathrm{2d}}(\gamma = 0.75) = 2.3 \pm 0.2$ 
The $\nu_{\mathrm{2d}}$ exponents in each case are identical within the bounds of error, in 
accord with the Harris Criterion. Although only one intersection appears 
for $\gamma = 0.3$ and $\gamma = 0.5$, the $\phi_{2}$ curves appear to intersect 
twice for $\gamma = 0.75$ with the second crossing near $\tilde{r} = 1.0$. 
The incipient upper mobility edge is mirrored in the data collapse, which for $\gamma = 0.75$ is 
defocused in the vicinity of $\tilde{r} = 1.0$, where the scaling of $Y_{2}$ is 
controlled by the critical point marking the upper mobility edge.

A similar treatment identifies mobility boundaries for 
$D = 3$, and the corresponding $\phi_{2}$ data collapses and intersections are 
shown in Fig.~\ref{fig:Fig3}.  For $\gamma < 2.5$, only a single intersection marks the  
boundary between localized states for $\tilde{r} < \tilde{r}_{c}$ and extended states 
for $\tilde{r} > \tilde{r}_{c}$. On the other hand, two sets of $\phi_{2}$ intersections may
be discerned for $\gamma > 2.5$,
indicating two distinct mobility edges with an interval of extended states 
flanked by localized states for $\tilde{r} < \tilde{r}_{c}^{\mathrm{lower}}$ and
$\tilde{r} > \tilde{r}_{c}^{\mathrm{upper}}$. 
In the context of a similar
tight binding model, a calculation by J.~J. Krich and A.~A-Guzik~\cite{Krich}
has also identified two sets of mobility edges for $D = 3$.  The case $\gamma = 2.5$
is marginal, since the upper intersection appears to 
coincide with the upper $\tilde{r}$ extreme, $\tilde{r} = 1$.
 The convergence of the mobility edges, and the 
concomitant constriction of the extended region affects the upper part of the data collapses
calculated for the lower mobility edges
where the severity of the defocusing increases with the proximity of the critical points in the upper mobility 
boundary.  
The areas in the right panels marked in gray match the $\tilde{r}$ domains 
between intersections of $\phi_{2}$ curves, and are determined from the extrapolated IPR results
in Fig.~\ref{fig:Fig1}, where we adopt the criterion $Y_{2} \leq 10^{-3}$ for the presence of 
extended states.
For the lower mobility edge, scaling collapses are sharp enough to be of service in fixing the 
exponent $\nu_{\mathrm{3d}}$, with results shown in Table~\ref{tab:nuvalues}.  As in the 2D case, 
within error bounds, the critical indices are 
in agreement across the broad range of $\gamma$ values under consideration.
\begin{table}
\begin{center}
\begin{tabular}{|c|c|c|c|c|c|}
\hline
$\gamma$ & 1.0  & 1.5  & 2.0 & 2.5 & 3.0 \\
\hline
$\nu_{\mathrm{3d}}$ & $1.1\pm 0.15$ & $1.2 \pm 0.15$ & $1.2\pm 0.15$ & $1.2\pm 0.15$ & $1.3 \pm 0.15$ \\
\hline
$\gamma$ & 3.5  & 4.0  & 4.5 & 5.0 & \\
\hline
$\nu_{\mathrm{3d}}$ & $1.2\pm 0.15$ & $1.2 \pm 0.15$ & $1.1\pm 0.15$ & $1.0\pm 0.2$ &  \\
\hline

\end{tabular}
\caption{\label{tab:nuvalues} $\nu_{\mathrm{3d}}$ results with for various $\gamma$ values}
\end{center}
\vspace{-0.6cm}
\end{table}

\begin{figure}
\includegraphics[width=.35\textwidth]{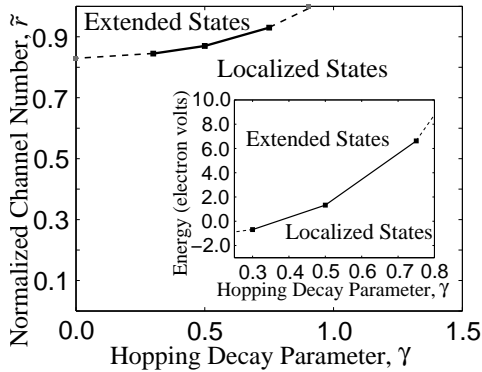}
\caption{\label{fig:Fig4} Phase portraits for $D = 2$ in terms 
of the normalized channel number $\tilde{r}$. The  
inset phase diagram is calculated with respect to energy.
Filled symbols are calculated, and broken lines are extrapolations.}
\end{figure}

\begin{figure}
\includegraphics[width=.45\textwidth]{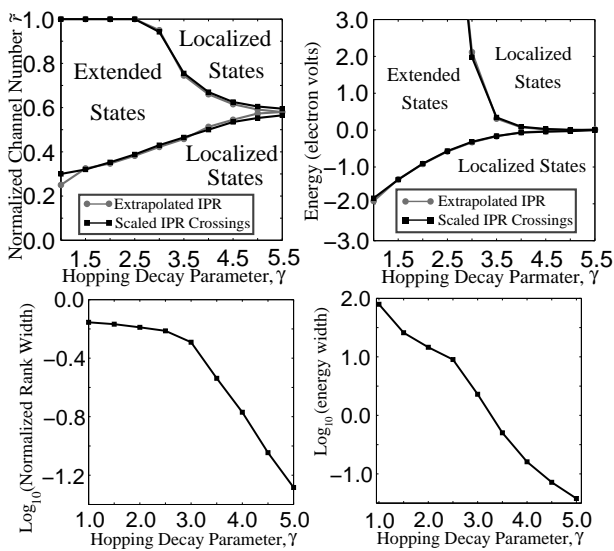}
\caption{\label{fig:Fig5} Phase portraits for $D = 3$ are shown in the 
upper row; mobility edges obtained from scaled IPR curve intersections 
and extrapolated participation ratios appear on the same graph.   
The logarithm of the width of the extended zone 
is graphed with respect to $\gamma$ in the lower panels}
\end{figure}

Phase diagrams showing regions of extended and localized states for $D = 2$ appear in 
Fig.~\ref{fig:Fig4}, where the main graph shows results in terms 
of the normalized rank $\tilde{r}$, and the inset is a phase portrait 
expressed in terms of the carrier state energies.
Extended states occupy the upper left corner of the phase diagram,
for $\gamma < 1.0$ and in the upper $E$ and $\tilde{r}$ ranges.
    
Phase portraits rendered in terms of $\tilde{r}$ and $E$ for the 3D case appear in the upper half of Fig.~\ref{fig:Fig5}; 
mobility edges calculated from $\phi_{2}$ intersections and 
gleaned from bulk $Y_{2}$ values appear in black and gray, respectively; the close overlap  
is a sign of excellent agreement among results of the two very distinct methods of 
locating mobility edges, and serves to validate the use of bulk $Y_{2}$ results in
constructing phase diagrams for 3D systems.  The lower part of Fig.~\ref{fig:Fig5}
contains graphs of the logarithm of the widths $w_{\tilde{r}}(\gamma)$ and $w_{\mathrm{E}}(\gamma)$ of the interval of 
extended states.  Although the extended region grows rapidly narrower for $\gamma \geq 3.0$, there is 
no indication of an abrupt termination for finite $l$; the widths, within the bounds of error,
 both are consistent with an asymptotically exponential scaling $e^{-A/l}$ for $l \ll 1$.

In conclusion, we have obtained phase diagrams for 2D and 3D amorphous systems with gas-like 
disorder, finding extended states to be supported in the upper part of 
the energy range for $D = 2$ above a threshold range 
$l_{c} \sim 0.9$
with the appearance of a second mobility edge near 
$l_{c}$.  On the other hand, for $D = 3$, we find a region of extended states even for 
$l \ll 1$ with no evidence for a termination of the extended phase for finite $l$.
Mobility edges are interpreted as lines of critical points, which we have characterized by 
calculating the corresponding critical exponents.

\begin{acknowledgments}
Useful discussions with E.~H. Hwang, J. Biddle, B. Wang, and S. Das Sarma
are acknowledged.
The numerical analysis has benefited from use of the 
University of Maryland, College Park HPCC  
computational facility.
\end{acknowledgments}



\begin{thebibliography}{11}
\expandafter\ifx\csname natexlab\endcsname\relax\def\natexlab#1{#1}\fi
\expandafter\ifx\csname bibnamefont\endcsname\relax
  \def\bibnamefont#1{#1}\fi
\expandafter\ifx\csname bibfnamefont\endcsname\relax
  \def\bibfnamefont#1{#1}\fi
\expandafter\ifx\csname citenamefont\endcsname\relax
  \def\citenamefont#1{#1}\fi
\expandafter\ifx\csname url\endcsname\relax
  \def\url#1{\texttt{#1}}\fi
\expandafter\ifx\csname urlprefix\endcsname\relax\def\urlprefix{URL }\fi
\providecommand{\bibinfo}[2]{#2}
\providecommand{\eprint}[2][]{\url{#2}}

\bibitem{Anderson} P.~W. Anderson, Phys. Rev. \textbf{109}, 1492 (1958).

\bibitem{Edwards} P.~P. Edwards and M.~J. Sienko, Phys. Rev. B \textbf{17}, 2575 (1978).

\bibitem{Dyson} F. J. Dyson, Phys. Rev. \textbf{92}, 1331 (1953).

\bibitem{Gade} R. Gade, Nucl. Phys. B \textbf{398}, 499 (1993).

\bibitem{Parshin}  D.~A. Parshin and H.~R. Schober, Phys. Rev. B \textbf{57}, 10232 (1998).

\bibitem{Cerovski} V.~Z. Cerovski, Phys. Rev. B \textbf{62}, 12775 (2000).

\bibitem{Takahashi} K. Takahashi and S. Iida, Phys. Rev. B \textbf{63}, 214201 (2001).

\bibitem{Logan1} D.~E. Logan and M.~D. Winn, J. Phys. C \textbf{21}, 5773 (1988).

\bibitem{Logan2} M.~D. Winn and D.~E. Logan, J. Phys.: Condens. Matter \textbf{1}, 1753 (1989).

\bibitem{Logan3} I.~J. Bush, D.~E. Logan, P.~A. Madden, and M.~D. Winn, J. Phys.:
Condens. Matter \textbf{1}, 2551 (1989).

\bibitem{Ching} W.~Y. Ching and D.~L. Huber, Phys. Rev. B \textbf{25}, 1096 (1982).

\bibitem{Blumen} A. Blumen, J.~P. Lemaistre, and I. Mathlouthi, J. Chem. Phys. \textbf{81},
4610 (1984).

\bibitem{Logan4} M.~K. Gibbons, D.~E. Logan, and P.~A. Madden, Phys. Rev. B \textbf{38}, 7292 (1988).

\bibitem{Krich} J.~J. Krich, A. Aspuru-Guzik, Phys. Rev. Lett. \textbf{106}, 156405 (2011).

\bibitem{Xiong} S.-J. Xiong and S.~N. Evangelou, Phys. Rev. B \textbf{64}, 113107 (2001).

\bibitem{Ziman} J.~M. Ziman, \textit{Models of Disorder:  The Theoretical Physics of 
Homogeneously Disordered Systems}, Cambridge University Press, P. 472, (1979).


\bibitem{priour} D.~J. Priour, Jr., cond-mat 1004.4366 (2010).

\bibitem{Santos} I.~F. dos Santos, F.~A.~B.~F. de Moura, M.~L. Lyra, and M.~D. Coutinho-Filho, J. Phys.: 
condens. Matter \textbf{19}, 476213 (2007).

\bibitem{Monthus} C. Monthus and T. Garel, Phys. Rev. B \textbf{81}, 224208 (2010).


\end{thebibliography}
\end{document}